\documentclass[11pt]{article}
\usepackage{moretext}
\usepackage{times}
\usepackage{mathfont}
\usepackage{graphicx}
\usepackage{cite}
\usepackage{url}
\def\min{\mathop{\rm min}}

\newtheorem{lemma}{Lemma}
\newtheorem{corollary}{Corollary}
\newtheorem{theorem}{Theorem}

\newcommand{\qed}{~$\Box$\medbreak}
\newenvironment{proof}{\noindent{\bf Proof: }}{\qed\medskip}

\DeclareSymbolFont{lasy}{U}{lasy}{m}{n}
\let\Box\undefined
\DeclareMathSymbol\Box{0}{lasy}{"32}

\DeclareSymbolFont{AMSb}{U}{msb}{m}{n}
\DeclareSymbolFontAlphabet{\Bbb}{AMSb}
\def\R{\ensuremath{{\Bbb R}}}
\def\O{\ensuremath{\mathcal{O}}}

\begin{document}
\bibliographystyle{abuser}

\title{Incremental and Decremental Maintenance of Planar Width}

\author{David Eppstein~\thanks{Dept. of
Information and Computer Science, Univ. of California, Irvine, CA
92697-3425, eppstein@ics.uci.edu. Work supported in part by NSF
grant CCR-9258355 and by matching funds from Xerox Corp.}}

\date{}
\maketitle
\thispagestyle{empty}
 
\begin{abstract}
We present an algorithm for maintaining the width of a planar point set
dynamically, as points are inserted or deleted.  Our algorithm takes
time
$\O(kn^{\epsilon})$ per update, where $k$ is the amount of change the
update causes in the convex hull, $n$ is the number of points in the set,
and $\epsilon>0$ is any arbitrarily small constant.  For
incremental or decremental update sequences, the amortized time per
update is $\O(n^{\epsilon})$.  
\end{abstract} 

\section{Introduction}

\begin{figure}[t]
$$\includegraphics[width=3in]{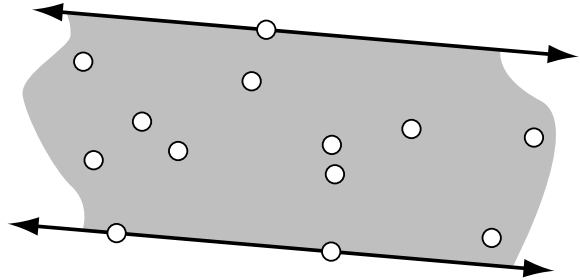}$$
\caption{The width of a point set is the minimum width of an infinite
strip containing the set.}
\label{fig:width}
\end{figure}

The {\em width} of a geometric object is the minimum distance between two
parallel supporting hyperplanes.  In the case of planar
objects, it is the width of the narrowest infinite strip that completely
contains the object (Figure~\ref{fig:width}).  The width of a planar point
set can be found from its convex hull by a simple linear time
``rotating calipers'' algorithm that sweeps through all possible slopes,
finding the points of tangency of the two supporting lines for each slope
\cite{PreSha-85,HouTou-PAMI-88,Tou-MELECON-83}.

Despite several attempts, no satisfactory data structure is known for
maintaining this fundamental geometric quantity {\em dynamically}, as
the point set undergoes insertions and deletions.
The methods of Janardan, Rote, Schwarz, and Snoeyink
\cite{Jan-IJCGA-93,RotSchSno-CCCG-93,Sch-EWCG-93} maintain only an
approximation to the true width.
The method of Agarwal and Sharir \cite{AgaSha-CGTA-91} solves only the
decision problem (is the width greater or less than a fixed bound?)
and requires the entire update sequence to be known in advance.
An algorithm of Agarwal et al. \cite{AgaGuiHer-WADS-97} can maintain the
exact width, but requires superlinear time per update
(however note that this algorithm allows the input points to have
continuous motions as well as discrete insertion and deletion events).
Finally, the author's previous paper \cite{Epp-CGTA-96} provides a fully
dynamic algorithm for the exact width, but one that is efficient only in
the average case, for random update sequences.

In this paper we present an algorithm for maintaining the exact width
dynamically.
Our algorithm takes time $\O(kn^{\epsilon})$ per
update, where $k$ is the amount of change the update causes in the convex
hull,
$n$ is the number of points in the set, and $\epsilon>0$ is any
arbitrarily small constant.  In particular, for {\em incremental}
updates (insertions only) or {\em decremental} updates (deletions only),
the total change to the convex hull can be at most linear and
the algorithm takes $\O(n^{\epsilon})$ amortized time per update.
For the randomized model of our previous paper, the expected value of
$k$ is $\O(1)$ and the average case time per update of our algorithm
is again $\O(n^{\epsilon})$.

Our approach is to define a set of objects (the features of the convex
hull), and a bivariate function on those objects (the distance between
parallel supporting lines), such that the width is the minimum value of
this function among all pairs of objects.  We could then use a data
structure of the author
\cite{Epp-DCG-95} for maintaining minima
of bivariate functions,
however in the case of the width this minimum is more easily maintained
directly.
To apply this approach, we need data structures for dynamic nearest
neighbor querying on subsets of features; we build these data structures
by combining binary search trees with a data structure of Agarwal and
Matou\v{s}ek for ray shooting in convex polyhedra \cite{AgaMat-Algo-95}.

\section{Corners and Sides}

\begin{figure}[t]
$$\includegraphics[width=3in]{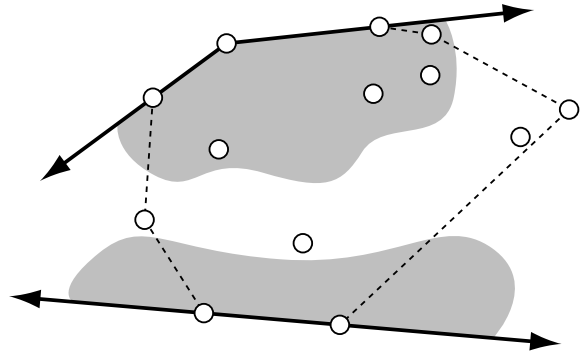}$$
\caption{A corner and an incompatible side of a point set.}
\label{fig:features}
\end{figure}

Given a planar point set $S$, we define a {\em corner} of $S$ to
be an infinite wedge, having its apex at a vertex of the convex hull of
$S$, and bounded by two rays through the hull edges incident to that
vertex.  We define a {\em side} of $S$ to be an infinite halfplane,
containing $S$, and bounded by a line through one of the hull edges.
Figure~\ref{fig:features} depicts a point set, its convex hull, a corner
(at the top of the figure), and a side (at the bottom of the figure).

We say a corner and a side are {\em compatible} if they could be
translated to be disjoint with one another, and {\em incompatible}
otherwise.  Alternatively, a side is compatible with a corner if the
boundary line of the side is parallel to a different line that is tangent
to the convex hull at the corner's apex.  The corner and side in the
figure are incompatible, because if one translates the side's boundary to
pass through the corner's apex, it would penetrate the convex hull.

Given a side $s$ and a compatible corner $c$, we define the {\em
distance} $d(s,c)$ to be simply the Euclidean distance between the
apex of the corner and the boundary line of the side.
Equivalently, this is the distance between parallel lines supporting the
convex hull and tangent at the two features. However, if $s$ and $c$ are
incompatible, we define their distance to be
$+\infty$.
Let ${\rm width}(S)$ denote the width of $S$,
${\rm sides}(S)$ denote the set of sides of $S$, and
${\rm corners}(S)$ denote the set of corners of $S$.

\begin{lemma}\label{lem:width-is-min}
For any point set $S$ in $\R^2$,
$${\rm width}(S) =
\min_{{\scriptstyle s\in {\rm sides}(S)}
\atop {\scriptstyle c\in {\rm corners}(S)}}
d(s,c).$$
\end{lemma}

\begin{proof}
Clearly, any compatible pair defines an infinite strip having width
equal to the distance between the pair, so the overall width can be at
most the minimum distance.
In the other direction, let $X$ be the infinite strip tangent on both
sides to the convex hull and defining the
width; then at least one of the tangencies must be to a convex hull
edge, for a strip tangent at two vertices could be rotated to become
narrower. The opposite tangency includes at least one convex hull
vertex, and the edge and opposite vertex form a compatible side-corner
pair.
\end{proof}

\begin{lemma}\label{lem:compatible}
Each side of the convex hull has at most two compatible corners.
The sides compatible to a given corner of the convex hull form a
contiguous sequence of the hull edges.
\end{lemma}

By Lemma~\ref{lem:compatible}, there are only $\O(n)$ compatible
side-corner pairs. The known static algorithms for width work by listing
all compatible pairs.
The dynamic algorithm of our previous paper maintained a graph,
the {\em rotating caliper graph}, describing all such pairs.
However such an approach can not work in our worst-case dynamic setting:
there exist simple incremental or decremental update sequences for which
the set of compatible pairs changes by $\Omega(n)$ pairs after each
update.  Instead we use more sophisticated data structures to quickly
identify the closest pair without keeping track of all pairs.
To do so, we will need to keep track of the set of convex hull features,
as the point set is updated.

\begin{lemma}[Overmars and van Leeuwen \cite{OveVL-JCSS-81}]
We can maintain a list of the vertices of the convex hull of a dynamic
point set in $\R^2$, and a data structure for performing logarithmic-time
binary searches in the list, in linear space and time $\O(\log^2 n)$ per
point insertion or deletion.
\end{lemma}

Recently, Chan \cite{Cha-FOCS-99} has improved these bounds to
near-logarithmic time, however this improvement does not make a
difference to our overall time bound.

\begin{lemma}\label{lem:track-features}
We can maintain a dynamic point set in $\R^2$, and keep track of its sets
of corners and edges, in linear space and time $\O(\log^2 n + k)$ per
update, where
$k$ denotes the total number of corners and edges inserted and deleted
as part of the update.
\end{lemma}

\begin{proof}
We apply the data structure of Overmars and van Leeuwen
from the previous lemma.  The
set of features inserted and deleted in each update can be found by a
single binary search to find one such feature, after which each adjacent
feature affected by the update can be found in constant time by
traversing the maintained list of hull vertices.
\end{proof}

\section{Finding the Nearest Feature}

In order to apply our closest pair data structure, we need to be able to
determine the nearest neighbor to each feature in a dynamic subset of
other features.  We first describe the easier case, finding the
nearest corner to a side.

\begin{lemma}\label{lem:near-corner}
We can maintain the corners of a point set in $\R^2$, and handle
queries asking for the nearest corner to a given side,
in time $\O(\log^2 n)$ per update and $\O(\log n)$ per query.
\end{lemma}

\begin{proof}
We use the same dynamic convex hull data structure as in
Lemma~\ref{lem:track-features}.  Each query can be answered by a single
binary search in the hull.
\end{proof}

We next describe how to perform dynamic nearest neighbor queries in the
other direction, from query corners to the nearest side.  To begin with,
we show how to find the nearest line to a corner, ignoring whether the
line belongs to a compatible side.

\begin{lemma}[Agarwal and
Matou\v{s}ek \cite{AgaMat-Algo-95}]\label{lem:AgaMat}
For any $\epsilon>0$,
we can maintain a dynamic set of halfspaces in $\R^3$, and
answer queries asking for the first halfspace boundary hit by a ray
originating within the intersection of the halfspaces, in time
$\O(n^{\epsilon})$ per insertion, deletion, or query.
\end{lemma}

\begin{lemma}\label{lem:rayshoot}
We can maintain a dynamic set of halfplanes in $\R^2$, and handle queries
asking for the nearest halfplane boundary to a given query point, where
the query is required to be in the intersection of the halfplanes, in time
$\O(n^\epsilon)$ per query, halfplane insertion, or halfplane deletion.
\end{lemma}

\begin{proof}
For a given halfplane $H$, let $D_H(x,y)$ denote $\pm 1$ times the
distance of point $(x,y)$ to the boundary of $H$, where the factor is
$+1$ for points in the halfplane and $-1$ for points outside the
halfplane.  $D_H$ is a linear function and can be used to define a
three-dimensional halfspace $\{(x,y,z) : D_H(x,y) \ge z\}$.
Then $D_H(x,y)$ is equal to the vertical distance from point $(x,y,0)$ to
the boundary of this halfspace.

Maintain such a three-dimensional halfspace for each of the halfplanes
in the set, along with the data structure of Lemma~\ref{lem:AgaMat}.
A nearest halfplane query from point $(x,y)$
can be answered by performing a vertical ray shooting query from point
$(x,y,0)$; the first halfspace boundary hit by this
ray corresponds to the nearest halfplane
to the query point.
\end{proof}

\begin{lemma}\label{lem:near-side}
We can maintain the sides of a point set in $\R^2$, and handle
queries asking for the nearest side to a given corner,
in amortized time $\O(n^\epsilon)$ per query, side insertion, or side
deletion.
\end{lemma}

\begin{proof}
We store the sides in a weight-balanced binary tree \cite{NieRei-SJC-73},
according to their positions in cyclic order around the convex hull.
For each node in the tree, we store the data structure of
Lemma~\ref{lem:rayshoot} for finding nearest boundaries among the sides
stored at descendants of that node.

For each query, we use the binary tree to represent the contiguous group
of compatible sides (as determined by Lemma~\ref{lem:compatible}) as the
set of descendants of $\O(\log n)$ tree nodes.  We perform the vertical
ray shooting queries of Lemma~\ref{lem:rayshoot} in the data structures
stored at each of these nodes, and take the nearest of the $\O(\log n)$
returned sides as the answer to our query.

Each update causes $\O(\log n)$ insertions and deletions to the data
structures stored at the nodes in the tree, and may also cause certain
nodes to become unbalanced, forcing the subtrees rooted at those nodes
to be rebuilt.  A rebuild operation on a subtree containing $m$ sides
takes time $\O(m^{1+\epsilon})$, and happens only after $\Omega(m)$
updates have been made in that subtree since the last rebuild, so the
amortized time per update is $\O(n^\epsilon)$.
\end{proof}

\section{Dynamic Width}

We are now ready to prove our main result.

\begin{theorem}
We can maintain the width of a planar point set in $\R^2$, as points are
inserted and deleted, in amortized time $\O(kn^\epsilon)$ per insertion or
deletion, where $k$ denotes the number of convex hull sides and corners
changed by an update.
\end{theorem}

\begin{proof}
We store the data structures described in the previous lemmas,
together with a pointer from each corner of the point set to the nearest
side (this pointer may be null if there is no side compatible to the
corner).  Finally, we store a priority queue of the corner-side pairs
represented by these pointers, prioritized by distance.
By Lemma~\ref{lem:width-is-min}, the minimum distance in this priority
queue must equal the overall width.

When an update causes a corner to be added to the set of features,
we can find its nearest side in time $\O(n^\epsilon)$ by
Lemma~\ref{lem:near-side}, and add the pair to the priority queue in
time $\O(\log n)$.

When an update causes a corner to be removed from the set of features,
we need only remove the corresponding priority queue entry,
in time $\O(\log n)$ per update.

When an update causes a side to be added to the set of features,
at most two corners can be compatible with it
(Lemma~\ref{lem:compatible}). We can find these compatible corners by
binary search in the dynamic convex hull data structure used to maintain
the set of features, in time $\O(\log n)$.  For each corner, we compare
the distances to the new side and the side previously stored in the
pointer for that corner, and if the new distance is smaller we change the
pointer and update the priority queue.

Finally, when an update causes a side to be removed,
that side can be pointed to by at most the two corners compatible with
it.  We use the dynamic convex hull data structure to find the
compatible corners, and if they point to the removed side, we recompute
their nearest side in time $\O(n^\epsilon)$ by Lemma~\ref{lem:near-side}.
\end{proof}

\begin{corollary}
We can maintain the width of a point set in $\R^2$ subject to insertions
only, or subject to deletions only, in amortized time $\O(n^\epsilon)$ per
update.
\end{corollary}

\begin{proof}
For the incremental version of the problem, each insertion
creates at most two new sides and three new corners, along with deleting
$h+2$ corners and $h+1$ sides where $h$ is the number of input points
that become hidden in the interior of the convex hull as a consequence
of the insertion.  Each point can only be hidden once, so the total
number of changes to the set of sides and corners over the course of the
algorithm is at most $10n$.  The argument for deletions is equivalent
under time-reversal symmetry to that for insertions.
\end{proof}

We note that in the average case model of our previous paper on dynamic
width \cite{Epp-CGTA-96}, the expected value of $k$ per update is
$\O(1)$, and therefore our algorithm takes expected time $\O(n^\epsilon)$
per update.  This is not an improvement on that paper's $\O(\log n)$
bound, but it is interesting that our algorithm here is versatile enough
to perform well simultaneously in the incremental, decremental, and
average cases.

\section{Conclusions and Open Problems}

We have presented an algorithm for maintaining the width of a dynamic
planar point set.  The algorithm can handle arbitrary sequences
of both insertions and deletions, and our analysis shows it to be
efficient for sequences of a single type of operation, whether
insertions or deletions.  Are there interesting classes of update
sequences other than the ones we have studied for which the total
amortized convex hull change is linear? Does there exist an efficient
fully dynamic algorithm for planar width?

Another question is to what extent our algorithm can be generalized to
higher dimensions.  The same idea of maintaining pairwise distances
between hull features seems to apply, but becomes more complicated.
In three dimensions, it is no longer the case that
incremental or decremental update sequences lead to linear bounds on the
total change to the convex hull, but it is still true that random update
sequences have constant expected change.  In order to apply our approach
to the three-dimensional width problem, we would need dynamic closest
pair data structures for finding the face nearest a given corner, the
corner nearest a given face, and the opposite edge nearest a given edge.
The overall expected time per update would then be $\O(\log^2 n)$ times
the time per operation in these closest pair data structures.
Can this approach be made to give an average-case dynamic
algorithm for three dimensional width that is as good as the best known
static algorithms
\cite{ChaEdeGui-DCG-93,AgaSha-DCG-96}?

\bibliography{dynwidth}
\end{document}